# Simple Relativity Approach to Special Relativity

Ralph Berger, PhD, UC Berkeley, Feb.13, 2020

Abstract: The development of both special and general relativity is accomplished in a series of 6 papers using a simple approach. The purpose is to explain the how and why of relativity to a broad public, and to be useful for students of physics by providing alternate ways to develop and view relativistic phenomena. In this first paper, the rules for special relativity are developed to explain velocity-related time dilation and length contraction, and the interchangeable nature of mass and energy. In the second paper, conservation of energy is applied to show how gravity affects time speed and fall velocities. In the third paper, the remainder of general relativity is developed based on gravitational impacts on length measurements, and the effect known as the Shapiro Time Delay is explained. In the fourth paper, these rules are used to explain the precession of satellites and planets. In the fifth paper, these rules are used to explain gravitational lensing. In the sixth paper, the appearance of Lorentz contraction is discussed along with a simple resolution of the Ehrenfest paradox.

Introduction: Simple Relativity is a redevelopment of Special and General Relativity using the approach that velocity and gravity affect time speed and length measurements. This paper develops the equations for ϒ, the time speed impact of velocity. Concepts are developed quickly here using familiar approaches to generate familiar results. The brevity will make this paper difficult for anyone not familiar with relativity; it is more of an overview that can be used as a description of the logical development of the concepts.

The one postulate required is that light speed c be the same to all observers. We come to this result from considering Maxwell's development of the speed of an electromagnetic wave as being calculable without a reference media, hence the velocity is a fixed value, but it is fixed without any reference frame. Einstein's leap was to conclude that it must be fixed for all reference frames, and the observer's sense of time speed must change to satisfy the measurement of lightspeed.

**Time Dilation: Velocity causes Time Speeds to Differ in Different Reference Frames**

The relative time speed due to velocity ϒ is found using the relation of length = velocity*time. A pole of length L can be measured by use of a photon's path by speed of light c * time traveled t. Placing this pole on a passing spaceship allows two observers to see the photon's path as having different lengths. The pilot measures the pole length by his time $t_o$ while the still observer sees a longer photon path ct due to the horizontal motion of the pass by velocity v multiplied by the still observer's time t. Of course, our observers will need to back calculate to address the time for the image to reach their eyes, but they can do that and we ignore any such necessary corrections for this development.

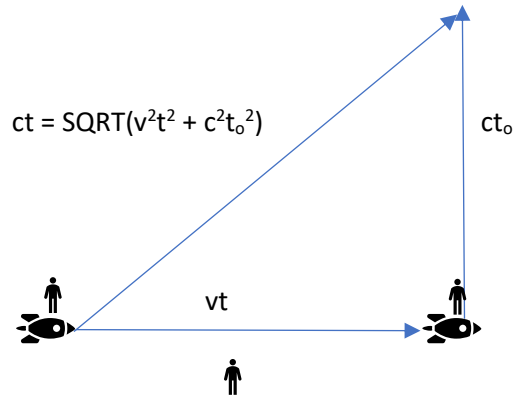

Figure 1: Lengths Measured by Velocity*Time

Figure 1 provides a relation in the length of the hypotenuse between the two observed times t and $t_o$. This is solved algebraically to obtain the relative time speed due to velocity commonly called the Lorentz factor[i].

$$\gamma = t/t_o = \frac{1}{\sqrt{1 - v^2/c^2}} \qquad [1]$$

**Nothing can Travel Faster than Light; Time Speed is Reduced (Never Increased) by Velocity**

Figure 1 also provides the additional insights that, no matter how great velocity v might grow, the length ct must be longer than vt. For the view of light speed to be the same for both observers, we must always have v<c. By the same measure, ct must always be larger than $ct_o$, implying that the observer can only calculate that the clock of the other is running slower than his own. This is a symmetric thing; the pole might be next to the still observer and the spaceship pilot see the longer photon path. That is, in special relativity, one always sees time move at its natural pace in one's own reference frame, but will see the time speed of others as slower (never faster) according to Equation 1.

**Measurements of Path Length are Reduced by Velocity**

In Figure 1, the horizontal displacement of vt is seen by the pilot to be $vt_o$, which is a shorter path length. Moving objects will see their path lengths shortened by the factor ϒ, hence this factor is also commonly called the Lorentz contraction. Lorentz contractions help explain why fast-moving travelers don't see themselves as exceeding lightspeed. For example, if traveling 1 lightyear at .866c, so that ϒ=2, we on Earth see the trip as requiring 1/.866 = 1.2 years, while the pilot records only 0.6 years. Traveling 1 lightyear in 0.6 years would seem to be faster than light, but due to the Lorentz contraction, the pilot will insist he traveled only 0.5 lightyears.

It is often said that the Lorentz contraction causes fast-moving objects to squish up in the calculations of still observers, so that a spaceship traveling such that ϒ=2 becomes half as long. That is useful in some calculations, such as the time for a spaceship to pass a particle point, but it is misleading in other situations. The traveler never feels squished, and the image of a fast-moving object is distorted by the time lag of photons that, in fact, make approaching objects look longer rather than shorter. That will be addressed in our sixth paper.

**Transforms and Metric Tensors**

If we square the right two terms of Equation 1 and rearrange algebraically, we can generate a simple rule for calculating the local time (often called proper time[ii]) for straight line travel in terms of the time and path length seen from a distance (sometimes called coordinate time). Here x = vt/c and is the distance traveled in units such as lightyears.

$$t_o = \sqrt{t^2 - x^2} \qquad [2]$$

For example, an astronaut traveling 5 years by Earth time, and going a distance of 4 lightyears by Earth measure, will experience a time of SQRT(25-16) = 3 years. Such conversions from one reference frame to another are call transforms, while the rules for how we measure things like time and distance are called metric tensors, or simply metrics. While they can become very complicated, in simple relativity we use only this one-dimensional form, since our purpose is to describe the how and why of relativistic impacts in general concepts and simple applications.

**Relativistic Doppler Effect Makes Time Speeds Appear Faster or Slower to the Viewer**

The apparent time speed of an approaching or receding object as measured in its image will be a product of ϒ and the conventional Doppler[iii] effect. If the source of a wave of frequency f moving at speed c is moving away from a viewer with velocity v, the wavelength c/f is lengthened by the distance moved in one period v*T which equals v/f. This gives a perceived wavelength based on a new frequency $f_n$ according to:

$$c/f_n = c/f + v/f \quad \text{or} \quad 1/f_n = \frac{(1+v/c)}{f} \quad \text{or} \quad f/f_n = (1 + v/c) \qquad [3]$$

*Had the light source below been non-moving, the wavelength would be c/f*

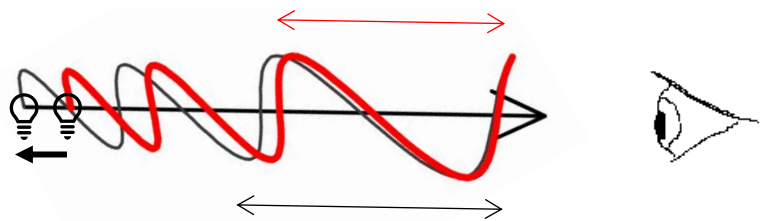

*Since the source moves v/f during each wave generation, the length becomes c/f + v/f*

Figure 2: Representation of Wavelength Increase by Motion, the Doppler Effect

The non-relativistic Doppler Effect of Equation 3 is modified for light sources to account for a further time change effect due to Equation 1. Recognizing that the frequency ratio is the same as the observed time ratio, we square Equation 3 and multiply it by ϒ² to get:

$$\left(t/t_n\right)^2 = \frac{(1+v/c)(1+v/c)}{(1+v/c)(1-v/c)} \quad \text{or} \quad \frac{t}{t_n} = \sqrt{\frac{(1+v/c)}{(1-v/c)}} \quad \text{on separation and} \quad \frac{t}{t_n} = \sqrt{\frac{(1-v/c)}{(1+v/c)}} \quad \text{on approach} \quad [4]$$

By measuring frequency changes of light spectra, which is the same as measuring relative observed time speeds, we can calculate the velocity v of approach or separation. We use this in astronomy to measure the expansion rate of the universe. That is, by measuring the changes in time speed, we can calculate velocity.

**Relativistic Velocity Addition**

We can also use the Doppler effect to explain relativistic velocity addition[iv]. If two twin astronauts leave Earth in opposite directions, they can calculate their relative velocity by means of a Skype session with their mother on Earth. Say Astronaut A leaves Earth at velocity u to the left and her twin brother leaves at velocity v to the right. In a Skype conversation, A will see her mother's video image slowed by Equation 4 applied to a separation velocity of u and her brother, on a screen in the background, slowed by an additional application of Equation 4 to a separation velocity of v. That is, Twin A calculates the speed difference u' from herself and her brother from:

$$\sqrt{\frac{1+u'/c}{1-u'/c}} = \sqrt{\frac{1+u/c}{1-u/c}} * \sqrt{\frac{1+v/c}{1-v/c}} \quad \text{which can be reduced with effort to} \quad u' = \frac{u+v}{1+uv/c^2} \quad [5]$$

The $c^2$ term can be removed from Equation 5 if u and v are expressed as fractions of light speed, that is, as numbers from 0 to 1. Equation 5 is tested by example in Table 1. Testing by example has the advantage of demonstrating that so long as u and v are less than 1, then u' must be less than one. This solves the apparent paradox that if one object moves at near light speed relative to a second, while the second moves at near light speed relative to a third, the relative speed of the first to the third is still less than light speed. To use the Skype analogy again, Twin A will see her brother in "double" slow motion on a screen behind her mother, but there will always be non-zero motion, so the relative velocity between A and her twin will always be less than light speed.

This last approach can be described mathematically as the time speed based on relative velocity u' is equal to the products of time speeds based on velocities u and v as:

$$ϒ_{u'} = ϒ_u * ϒ_v * (1+uv/c^2) \quad [6]$$

Equation 6 is a little harder to develop algebraically than Equation 5, but it can be tested by example.

Table 1: Demonstrations of Equations 5 and 6

| u/c | v/c | u'/c by Eq [5] | ϒ_u | ϒ_v | ϒ_u' by [6] | ϒ_u' by [1] |
|-----|-----|----------------|----------|----------|-------------|-------------|
| 0.7 | 0.9 | 0.981595 | 1.40028 | 2.294157 | 5.236314 | 5.236314 |
| 0.9 | 0.9 | 0.994475 | 2.294157 | 2.294157 | 9.526316 | 9.526316 |
| 0.5 | 0.4 | 0.75 | 1.154701 | 1.091089 | 1.511858 | 1.511858 |
| 0.2 | 0.99 | 0.993322 | 1.020621 | 7.088812 | 8.667516 | 8.667516 |
| 0.995 | 0.999 | 0.999997 | 10.01252 | 22.36627 | 446.5431 | 446.5431 |

**Relativistic Mass Increase**

Although problematic in some ways, we can use the once common concept of relativistic mass to define apparent mass increase with velocity. A common development is to image a particle moving, left to right, at speed v greater than half light speed, that suddenly divides into equal halves. In Figure 3, we use the reference frame of the left side half to state that the initial momentum is 2m$_v$v, and the momentum after division is m$_o$*0 + m$_u$*u. The problem with the momentum balance, 2m$_v$v = m$_u$*u, is that for m$_u$ to be half m$_v$ would require that v be twice u, and that would violate the idea that nothing can go faster than light.

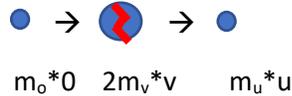

m$_o$*0    2m$_v$*v    m$_u$*u

Figure 3: Conservation of Momentum Applied to a dividing particle from the Reference Frame of the Left Half (in a time history, the particle is 2m$_v$ initially, and then divides into m$_o$ and m$_u$)

The solution to calculate u is to apply Equation 5, using the separation velocity of v for both velocities on the right, that is,

$$u = \frac{v+v}{1+v^2/c^2} = \frac{2v}{1+v^2/c^2}$$

Then the momentum balance and mass balances become:

$$2m_v v = \frac{2 m_u v}{1+v^2/c^2} \quad \text{and} \quad m_o + m_u = 2m_v \quad \text{so} \quad m_v = \frac{2m_v - m_o}{1+v^2/c^2}$$

With some algebraic manipulation, this becomes

$$m_v = \frac{m_o}{1-v^2/c^2} = \gamma_v^2 \, m_o \qquad and \qquad m_u = \frac{m_o}{\sqrt{1-u^2/c^2}} = \gamma_u \, m_o \qquad [7]$$

The right-hand relation is the one used for "relativistic mass," while the left-hand relation is discussed below Table 2. Both are necessary for mass and momentum balances to work, as seen in Table 2 examples, where velocities are expressed as fraction of light speed, the mass balance is 2m$_v$ = m$_o$+m$_u$ which we see is really 2$\gamma_v^2$ m$_o$ = m$_o$ + $\gamma_u$ m$_o$, and the momentum balance is the final two columns. The first row would be at low enough velocity (<1 million mph) for Newton's laws to work.

Table 2: Particle Division Examples using Equation 7

| m$_o$ | v | u by [5] | m$_v$ = $\gamma_v^2$m$_o$ | m$_u$ = $\gamma_u$m$_o$ | 2m$_v$ =m$_o$+mu | 2m$_v$v | m$_u$u |
|---|---|---|---|---|---|---|---|
| 1 | .001 | .002 | 1 | 1 | 2 | .002 | .002 |
| 1 | 0.1 | 0.19802 | 1.010101 | 1.020202 | 2.020202 | 0.20202 | 0.20202 |
| 1 | 0.4 | 0.689655 | 1.190476 | 1.380952 | 2.380952 | 0.952381 | 0.952381 |
| 1 | 0.8 | 0.97561 | 2.777778 | 4.555556 | 5.555556 | 4.444444 | 4.444444 |
| 1 | 0.9 | 0.994475 | 5.263158 | 9.526316 | 10.52632 | 9.473684 | 9.473684 |
| 1 | 0.99 | 0.999949 | 50.25126 | 99.50251 | 100.5025 | 99.49749 | 99.49749 |

**Mass Creation and Destruction, and Atomic Energy**

As developed in Table 2, the increase from $m_o$ to $m_v$ is by $\Upsilon_v^2$ yet the increase from $m_v$ to $m_u$ is by $\Upsilon_u$. The reason is remarkably that the mass of the unexploded particle, which is $m_v$ in Figure 3, contains a certain amount of mass-equivalent energy, which is the source of the energy that propels the two halves apart. An alternative description of the mass balance of this event written from the original particle's reference frame is $2m + 2(\Upsilon-1)m = \Upsilon m + \Upsilon m$, where the first term is the initial object's rest mass, the second the mass equivalent of the energy needed to cause the division, and the two terms on the right are the two created halves moving at high speed.

Written in that fashion, we are adding energy to create mass. The reverse direction would be akin to the fusion energy that powers the sun, that is, two particles each of mass $\Upsilon m$ fuse together into a heavier particle of mass $2m$ releasing the energy of $2(\Upsilon-1)m$.

In nuclear chemistry, we generally add up the masses of the reactants and the masses of the products, and determine that the energy needed or released is proportional to the mass created or destroyed. This is the energy of atomic bombs and nuclear power plants. The constant of proportionality between E and the created or destroyed mass m is $c^2$.

**E=mc²**

Our development of $E=mc^2$ works by postulating that the energy put into accelerating a mass will equal the creation of relativistic mass over and above rest mass[v]. That is, where F is the force creating the acceleration and energy by definition is the integral of force * distance, we want to prove that:

$$E = \int_0^s F\,ds = (m - m_o)c^2 = (\Upsilon - 1)m_o c^2 \qquad [8]$$

In the final term, we used Equation 7, $m = \Upsilon m_o$. This integral is solved by expressing force in terms of distance. We start with Newton's law again using the right-hand side of Equation 7[vi]:

$$F = \frac{d(mv)}{dt} = \frac{d}{dt}\frac{m_o v}{\sqrt{1-v^2/c^2}} = \frac{d}{dt}m_o v(1 - v^2/c^2)^{-\frac{1}{2}} = \frac{m_o a}{(1-v^2/c^2)^{3/2}} = \gamma^3 m_o a \qquad [9]$$

The steps to reach the final relation in Equation 9 are compressed, but it is a straight-forward derivative by parts with a substitution of $a = dv/dt$. We then apply conservation of energy to state that the acceleration history is not important to the total energy, and we can simplify the integral by assuming a constant acceleration. This both makes "a" a constant and $v^2 = 2sa$ by Newton's laws.

Then we can integrate to obtain:

$$E = \int_0^s m_o a(1 - 2sa/c^2)^{-\frac{3}{2}}ds = m_o c^2(1 - 2sa/c^2)^{-\frac{1}{2}} - m_o c^2 = (\gamma - 1)m_o c^2 \qquad [10]$$

As desired. Again, the math is compressed, but the student can work through it, solving the integral and putting in the limits of s and 0. The final step involves resubstituting $v^2 = 2sa$. $E=(\Upsilon-1)m_o c^2$ is properly the relation between using energy to create mass (as stars do when creating heavier elements) or the energy available from destroying mass (as fission reactors do when splitting atoms). The form $E=m_o c^2$ is the energy available should all the available mass be transformed, something that, as far as we know, only occurs at the subatomic level. But in theory, all the mass of an object could be transformed to this energy, which is often referred to as the rest mass energy, and then reconverted back into mass.

**Relativistic Kinetic Energy**

Looking at the above development, we see that Equation 10 defines the energy associated with motion, the kinetic energy[vii], as:

$$KE = (\Upsilon - 1)m_o c^2 \qquad [11]$$

This initially appears very different from the classical $KE = \frac{1}{2} mv^2$, but in fact, $\Upsilon$, which is $(1-v^2/c^2)^{-1/2}$ can be expanded by the binomial series to $1 + 1/2\, v^2/c^2 + 3/8\, v^4/c^4 + 15/48 * v^6/c^6 + \ldots$ For small values of v/c, only the first two terms are significant. Substituting them in for $\Upsilon$ gives the classical form of KE.

**Applications**

Using the simple relations developed here allows for simple solutions to special relativity paradoxes. For example, the twin astronaut paradox is the case in which one twin travels rapidly from Earth and then comes back while her brother stays at home. In the eyes of each, they are still while the other recedes, and then returns. Why is it that the Astronaut twin's clock records less time than the brother's clock? This is puzzling if we accept too strongly the contention that Equation 1 is symmetric for each observer.

The first simple relativity solution is to use the Doppler effect of Equation 4. If the twins keep their eyes on each other through magic telescopes, they will indeed see equal slowed-down aging while separating, and equal sped-up aging while approaching. However, the astronaut twin sees her brother begin the sped-up period immediately on turning around. Due to a time lag because of the speed of light, the brother doesn't see her sister turn around until a longer period has passed. Hence, relative to his sister, the Earth brother sees a longer period of slowed-down aging and a briefer period of sped-up aging. When comparing clocks after the journey, both agree that the brother's clock was seen to have recorded more time. The discrepancy is due to the sister's return leg, in which she is "flying upstream" against the flow of photons carrying the image of the brother.

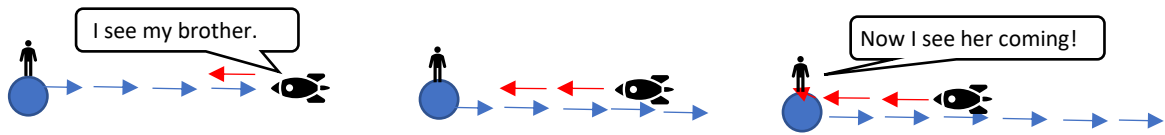

Figure 4: Returning astronaut sees Earth before Earth can see her return

The second simple relativity solution is to use a third observer at any arbitrary non-accelerating reference frame. The third observer will see the brother's progress as a constant velocity*time, whereas the sister might have less velocity for one leg, she will need to make up and have greater velocity for the other leg. The third observer can use Equation 1 and decree that the sister had a period of higher velocity which overall will result in slower time, in exactly the ratio needed to explain the final clock comparison. The discrepancy is due to the need of the sister to cover a greater distance as seen by a third observer, and hence the need to have a greater velocity. This third-observer explanation also explains why geostationary clocks (in satellites whose 24-hour orbit keeps them above a fixed point on Earth's equator) have a slower velocity component of time speed than earth-bound clocks, even though there is a zero velocity between the two clocks.

The relative pass by time paradox involves two identical spaceships of length L passing in opposite directions. Say the pilot of each is in a forward cockpit, and the first pilot records the time between tip passing and tail passing as t. She concludes that the other pilot has a slower running clock, so his time for the event is $t_o$, which by Equation 1 is a shorter time by $t/\Upsilon$. The second pilot, however, performs the same measurement and finds the value of t, while assuming the first pilot's time must be a shorter value $t_o$. Who recorded the larger time?

The simple relativity explanation is that both pilots do, indeed, record times of t for their experience of tip to tail and see times of $t_o$ recorded on the other's wristwatch. However, their view of the pass by involves seeing the other ship as shortened by a factor of $\Upsilon$ due to Lorentz contraction. Hence for the tip to pass by them, the travel distance is $L/\Upsilon$. But the other pilot will require the full distance L to reach the end of their ship, which increases the time requirement to be greater by a factor of $\Upsilon$. That factor gets canceled out by the slower running clock, hence both pilots record time t and expect the other pilot to record time t for their tip to tail trip. However, each pilot thinks that their ship is longer than the other's ship.

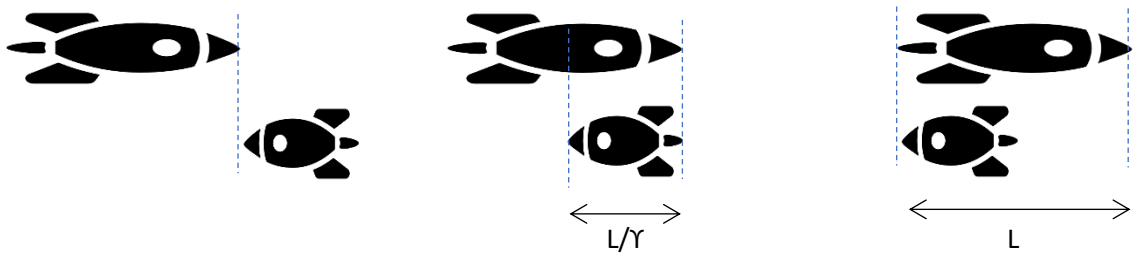

Figure 5: Each Pilot sees the Pass By Distance as L/Υ, but assumes the other Pilot sees it as L

The ladder in a barn paradox is that while a 30' ladder cannot fit inside a 20' barn when still, if it is shooting through opposite-side open barn doors with $\Upsilon=2$, there should be a time when a barn-based observer sees the ladder as 15' long and sees it as fitting inside, so quick door closing should trap it.

The simple relativity solution is to repeat that nothing really squishes due to velocity, but rather path lengths change. Yes, the time recorded at the exit door will be based on the ladder length of 15' divided by velocity, but the ladder never really is 15' long. It always stays 30 feet and is never trappable. If a farmhand closes the exit door just before the tip arrives, his view of the ladder is of it as still sticking out the entrance (since the image takes time to arrive). If the tip is stopped by the closed exit door, the signal that the tip has stopped is transmitted through the ladder at no faster than light speed. The ladder collapses like an accordion from tip down towards tail. Meanwhile, the tail of the ladder is still moving with no possible way of knowing that the front end has stopped. By the time the tail discovers that the tip has stopped, it is indeed inside the barn (and a second farmhand can close the entrance door), but the ladder has been physically squished by the collision and is now less than 20 feet long in reality. We return to this paradox in more detail in our 6th paper.

Conclusion

This paper has developed the relations for special relativity using conventional models, but has not used non-Newtonian concepts such as Minkowski diagrams or time-as-a-dimension. Its purpose is to set up an approach in which more difficult general relativity relations can be similarly developed. Subsequent papers in this series will develop this idea that all of relativity, special and general, can be explained in a prosaic fashion without difficult math or abstract concepts. Such an approach will have the advantage of both making relativity more available to an interested public, and provide an alternative simpler way of viewing the universe for students of physics. It will not produce any different results from the conventional approach to relativity, but it will provide the same results using much simpler math and with simple conceptual justification.

---

[i] Forshaw, Jeffrey; Smith, Gavin (2014). *Dynamics and Relativity*. John Wiley & Sons. ISBN 978-1-118-93329-9

[ii] Minkowski, Hermann (1908), "Die Grundgleichungen für die elektromagnetischen Vorgänge in bewegten Körpern", Nachrichten von der Königlichen Gesellschaft der Wissenschaften und der Georg-August-Universität zu Göttingen, Göttingen

[iii] Possel, Markus (2017). "Waves, motion and frequency: the Doppler effect". Einstein Online, Vol. 5. Max Planck Institute for Gravitational Physics, Potsdam, Germany

[iv] Relativistic velocity addition can be found in many places including https://en.wikipedia.org/wiki/Velocity-addition_formula

[v] There are many alternate ways of developing E=mc$^2$ including several listed at https://en.wikipedia.org/wiki/Mass%E2%80%93energy_equivalence

[vi] The factor of ϒ$^3$ in the force equation was noted by Einstein in his 1905 paper, On the Electrodynamics of Moving Bodies. He notes that we can retain the equation F=ma for motion in the direction of force and velocity if we use a "longitudinal mass" of ϒ$^3$m.

[vii] This definition of kinetic energy can be found in many places including https://en.wikipedia.org/wiki/Kinetic_energy#Relativistic_kinetic_energy_of_rigid_bodies